\documentstyle[aps]{revtex}
\tightenlines
\begin{document}
\title{Scaling for vibrational modes of fractals tethered at the boundaries}
\author{Sonali Mukherjee\cite{present} and Hisao Nakanishi} 
\address{Department of Physics, Purdue University, W. Lafayette, IN 47907}
\date{\today}
\maketitle
\begin{abstract}
Using a recently introduced mapping between a scalar elastic network
tethered at its boundaries and a diffusion problem with permanent traps,
we study various vibrational properties of progressively tethered
disordered fractals. Different scaling forms are proposed for different
types of boundary tethering and numerical investigation of the scaling
is performed by the approximate diagonalization of the corresponding large,
sparse transition probability matrices. Rather different localization
behaviors are found for the leading modes depending on the types
of tethering.
\end{abstract}

\newpage 
\section{Introduction}
\label{sec:Intro}

In this work we consider the vibrational spectrum associated with the
{\em scalar} elasticity of fractals when their boundaries are
progressively tethered to immobile, external anchors. By scalar elasticity,
we mean the vibrational problem where the local displacement variable $u_i (t)$
at site $i$ is a scalar. We study the various signatures of the vibrational
density of states and the localization properties of the normal modes
under these conditions.

The spatial scale invariance of fractals \cite{feder} and the absence
of translational invariance due to the boundaries conspire to influence
their vibrational spectra in fundamental ways. The now well-studied
fracton \cite{alexander} spectrum, with the characteristic,
fractional power law for the low-energy density of states,
arises for the bulk fractals with free boundaries, while so-called
fractino \cite{sapoval} spectrum arises for objects which do not have
fractal bulk but have fractal boundaries that are clamped. Here, we
generalize these further to study bulk fractals whose fractal boundaries
are progressively clamped or tethered to immobile anchors. The resulting
vibrational spectrum is complex yet shows some simple scaling features
common to many phenomena with long range order.

The problem of vibrations in a disordered system is in itself of physical
interest.  Granular systems \cite{herrmann}, such as sand or snow piles
or crops stored in silos, are among the systems of potential application.
For example, sound propagation in a granular system had been
modeled by Leibig \cite{leibig} using a scalar elastic network with
a random distribution of spring constants and analyzed in terms of
the normal modes. Boundary conditions such as tethering add much complexity
to the problem which may be relevant to some physical and engineering
applications. For example, random mixtures of soft and hard materials
such as solid granules embedded in a polymer gel or colloidal particles 
and aggregates filling the pores or cracks in rocks may constitute such
an application.  Another potential interest might be due to the recently
discovered anomalous behavior \cite{helium} of the superfluid transition
when aerogels \cite{aerogel} are embedded in liquid $^4$He or $^3$He,
where the aerogels may impose a boundary condition over a fractal boundary
of the liquid helium.

Most previous works on the effect of clamped boundaries on the vibrational
density of states of inhomogeneous systems dealt with deterministic 
inhomogeneities (such as in {\em fractal drums} \cite{sapoval}). In this
case the bulk is nonfractal (Euclidean) and the boundary is an ordered fractal,
and the effect is a correction term to the leading behavior, the latter
remaining the same as for a homogeneous system as one approaches the
asymptotic limit of large system sizes. We have a substantially different
system where both the bulk and the tethered boundaries are statistical 
fractals, sometimes of different fractal dimensionality, where even
the leading behavior is expected to be special to inhomogeneous systems.

The prototype of such a system is a critical percolation cluster
\cite{stauffer}. We have used site percolation clusters
created near the critical percolation threshold on square and
simple cubic lattices.  We focus our attention on the low-energy regime
where we expect the clamping effect of the fractal boundaries
to be pronounced. Without tethering, these modes tend to have
relatively large spatial extent, and thus are more sensitive
to boundary tethering. In particular, we study the lowest energy mode
($\epsilon_1$) and the peak ($\epsilon_p$) that appears in the low-energy
region of the vibrational density of states. 
The lowest energy mode is computationally
much easier to obtain than the entire density of states and yet captures
some of its essential features.

To study these systems, 
we exploit the mapping between scalar elasticity and diffusion as
extended \cite{nagoya} to the case of mapping between 
the vibration of tethered objects and diffusion with permanent traps.
In this approach, we define a transition probability matrix ${\bf W}$
(an $S\times S$ matrix for a cluster of $S$ sites) where
$W_{ij}$ is the hopping probability per step $p_{ij}$ from site $j$ to
site $i$ in the corresponding diffusion problem. We set $p_{ij}=1/z$ for
all pairs of nontethered sites $i,j$ where $z$ is the lattice coordination
number. The diagonal elements $W_{ii}=0$ if site $i$ is tethered
(representing the complete leakage of diffusion field or {\em full} tethering)
while, if $i$ is not tethered, $W_{ii}=1-n_i/z$ (representing 
conservation of diffusion field) where $n_i$ is the number of available
neighbors for site $i$. Then the time evolution of the diffusion field
$P_i (t)$ is given by
\begin{equation}
P_i (t+1) = \overline{\sum_j} [p_{ij} P_j (t) + (\delta_{ij}- p_{ji}) P_i(t) ]
          = \overline{\sum_j} W_{ij} P_j (t) \;,
\label{eq:evolution}
\end{equation}
where $\overline{\sum}$ includes the diagonal terms.

The eigenmodes of ${\bf W}$ are the normal modes of the tethered
vibration problem and the eigenvalues $\lambda$ are related to the classical
vibration frequency $\omega$ and energy $\epsilon \equiv \omega^2$ by
\begin{equation}
\lambda = 1 - \omega ^2 \approx \exp (-\omega ^2 )
\label{eq:eigen}
\end{equation}
where the last approximation corresponds to the long time limit.
We use this formulation in part to take advantage of the fact that the
low-energy region of the spectrum corresponds to the region of the
maximum eigenvalues of ${\bf W}$ \cite{review} which affords a much
easier access numerically. The numerical technique used is based on
the work of Saad \cite{saad} which implemented Arnoldi's method.
Preliminary results of this work were reported earlier in \cite{nagoya}
which introduced the mapping as well as presented preliminary numerical
results only for the {\em all boundary} tethered case in two dimensions.
In the current work, we have improved data in both two and three dimensions
as well as for {\em hull} tethering (see below).

\section{Scaling of lowest energy mode for hull versus all boundary tethering}
\label{maxscaling}

In a cluster with tethered boundaries the lowest energy mode
$\epsilon_1$ crucially depends on how many and which sites are tethered.
Clearly, for an individual untethered mode, tethering of the
sites where the mode has a large component greatly affects the mode.
Statistically, an important classification of the boundary sites is
between the internal and external boundaries (or {\em hull}) \cite{hull}.
For a fractal like the critical percolation cluster, they are both fractal
objects with similar (but not identical) fractal dimensions; in two
dimensions, the hull has a fractal dimension $d_f^h=1.75$ 
while the internal boundaries have a fractal dimension $d_f^i=91/48$
(same as the bulk fractal itself and also as the fractal dimension of all
boundary $d_f^b$). However, the behavior of $\epsilon_1$
(and the density of states) is unexpectedly different depending on
if only (some of) the hull sites are tethered or (some of) both types of
boundary sites (i.e., the hull and internal boundary) are tethered,
as functions of the number $N$ of the tethered sites and the size $S$ of
the cluster, the two main parameters in these situations.

In order to understand the interplay of the different length scales we
write the scaling variable as the ratio of the relevant length scales
for the two cases.  The common length scale for both cases is the
average distance $l$ between the tethered sites where
\begin{equation}
l \sim (S/N)^{1/d_f} .
\end{equation}
We also extend the meaning of $N$ by $N \equiv f N_T$ where $f$ is the
fraction of the $N_T$ boundary sites which are tethered: $N_T=N_h$
(number of hull sites) for hull tethering, $N_T=N_a$ (number of all
boundary sites) for all boundary tethering. In particular,
this means $N$ can be smaller than one if $f$ is sufficiently small,
which indicates many random samples in the ensemble will have no tethers.

The second relevant length scale $L$ for hull tethering for the critical
percolation cluster is the the average diameter of the cluster itself,
\begin{equation}
L_c \sim S^{1/d_f}
\end{equation}
since the hull forms a connected set of external boundaries to whose
interior the normal modes are confined. In particular, the spatial extent
of the lowest energy mode is dictated by that of the interior sites at the
scale of the cluster diameter $L_c$). 
See the illustration of these lengths in Fig.1.

On the other hand, in the case of all boundary tethering, the second
relevant length $L_b$ is the average diameter of a {\em blob} in the cluster.
As sites of both hull and internal boundary are tethered in {\em all} boundary 
tethering, a normal mode is confined to sites which are internal to both
boundaries. i.e., in the blobs.
The blobs in this context may be the areas of the cluster with 
essentially no internal holes (empty sites in the interior), which are
connected together by links of less connectivity to form the overall cluster.
See the illustration of these lengths in Fig.1.
Though we do not have a precise characterization of a {\em blob},
it is clear that this sense of a blob is different from those
discussed earlier, e.g., in the sense of a merely multiply connected
component \cite{coniglio}.

Since one would expect these new blobs to go critical at the 
ordinary critical percolation threshold $p_c$, its diameter $L_b$
may be expected to follow a power law at $p_c$,
\begin{equation}
L_b \sim S^{y}
\end{equation}
for an appropriate universal exponent $y$ in the scaling regime of large $S$.
We note that both a precise identification of such blobs and the relation
between $L_b$ and $S$ are still open questions \cite{cuansing}. 

Let us consider {\em scaling} for the hull tethering case first.
In the limit of no tethering, the problem is reduced to that of so-called
{\em ants} \cite{review} where the lowest energy mode is the
{\em stationary} mode with $\epsilon =0$ for any $S$. On the other hand,
for asymptotically large $S$ and fixed $L_c/l >0$ (i.e., fixed $N >0$
and $S \rightarrow \infty$), the number of tethered sites $N$ becomes
negligible compared to cluster size $S$. Thus, the lowest energy mode
is arbitrarily close to stationarity and should behave essentially in
the same way as the second lowest energy mode (lowest energy, nontrivial
mode) of the nontethered case, i.e., as $\epsilon_1 \sim S^{-d_w/d_f}$
($d_w$ is the {\em walk dimension} and $d_f$ is the fractal dimension of
the cluster).  Thus, the scaling form of $\epsilon_1$ is expected to be
\begin{equation}
\epsilon_1 \sim S^{-d_w/d_f} F(L_c /l) = S^{-d_w/d_f} G(N) .
\end{equation}

Since $\epsilon_1 \rightarrow 0$ as $N \rightarrow 0$ with fixed $S$,
we must have $G(z) \rightarrow 0$ as $z \rightarrow 0$. If we assume
analyticity of $G(z)$ for small $z$, in the absence of obvious symmetry
requirements which might remove the linear term, it seems reasonable
to conclude $G(z) \sim z$ for small $z$.

It is possible to obtain the limit of $z \rightarrow \infty$ in such a 
way as to essentially remove the effect of tethering simultaneously.
To this end, consider joining many mini-clusters with a given individual
$l_1$ in such a way so as to make the overall cluster a fractal with
fractal dimension $d_f$.  In this process the overall cluster length scale
$L_c$ increases at a greater rate ($L_c \sim S^{1/d_f}$)
than the overall $l$ ($l \sim (S/N)^{1/d_f}$) (since $N$ is also growing).
Thus, as more mini-clusters are joined, the variable $L_c /l$ tends to
$\infty$ while the tethered hull ($S^{d_f^h/d_f}$) becomes a negligible
part of the entire boundary which goes as $S$ (since $d_f^h < d_f$).
In this case with the diminished importance of tethered hull, 
the behavior of $\epsilon_1$ should converge toward the lowest nontrivial
mode of the nontethered limit once again ($\epsilon_1 \sim S^{-d_w/d_f}$).  
Thus, in this limit of $z = N \rightarrow \infty$, the lowest eigenmode
is confined {\em entirely} to the interior of the cluster, and
the spatial extent and the energy of the mode is unaffected by further
tethering of hull sites. This is equivalent to {\em herding} \cite{nagoya}
which is a term we had used to describe the saturation effect of the
confinement of an eigenmode.  Thus, we must have
$G(z) \rightarrow const.$ as $N \rightarrow \infty$.

Next we consider all boundary tethering.
The no-tethering limit is of course the same as for the hull tethering case.
As one tethers sites of all boundaries, the lowest-energy mode gets
localized in blobs which are unaffected by tethering. 
Moreover, if the ratio $L_b/l$ is fixed at a nonzero value and
$S \rightarrow \infty$, then each {\em blob} behaves like a cluster of its
own with hull tethering. The expected scaling variable is then the number
of tethered sites per blob $N L_b^{d_f}/S$ (which is equal to $(L_b/l)^{d_f}$
as expected). We thus propose a scaling form
\begin{equation}
\epsilon_1 \sim S^{-d_w/d_f} \overline{F}(L_b /l)
= S^{-d_w/d_f} \overline{G}(N/S^x) ,
\label{eq:maxall}
\end{equation}
where $x=1-y d_f$.

For $z \rightarrow 0$, as for the hull tethering,
we have $\overline{G}(z) \rightarrow 0$. However, the limit of
$z \rightarrow \infty$ can be achieved by increasing the fraction
of tethered boundary sites as the cluster size increases. 
The limit should then correspond to the so-called {\em ideal chain}
\cite{review} with all the boundary sites tethered. 
Thus, in this limit, $N \sim S$ and $\epsilon_1 \sim (\ln S)^{-2/d_0 }$
\cite{slowest} (where $d_0$ is the exponent describing the stretched
exponential behavior of the density of states for ideal chains introduced
in \cite{achille}). Numerically, $d_0$ is about 4 in two dimensions and
about 6 in three dimensions for the critical percolation cluster.
Thus, the scaling function
$\overline{G}(z) \sim z^{d_w/[d_f (1-x)]}(\ln z)^{-2/d_0}$ as
$z \rightarrow \infty$.

It is interesting to consider the connection of this behavior to the
confinement of the lowest energy mode to the small, compact regions
of the cluster in this case. These regions are not affected by tethering
because of the absence of internal boundaries (holes). Their fractal
dimension must be essentially equal to the Euclidean lattice dimension $d$. 
Thus, we denote their average size by $S_c$, we may expect 
$\epsilon_1 \sim {S_c}^{-2/d}$ where 2 is the walk dimension on a compact
cluster. In view of the relation $\epsilon_1 \sim (\ln S)^{-2/d_0 }$ above,
this would lead to $S_c \sim (\ln S)^{d/d_0} \approx (\ln S)^{1/2}$ in
both two and three dimensions. This type of slow growth of $S_c$ is consistent
with our direct observations.

Fig.2 and Fig.3 show the numerical results for the above discussed
scaling behavior for the square lattice in two dimensions. Fig.2 is
for the hull tethering case while Fig.3 is for all boundary tethering.
Besides the scaling variables on the $x$ axis being different,
the scaling functions exhibit dramatically different behavior.
Though, in both cases, $\epsilon_1$ (and thus the scaling functions)
approach zero as $z \rightarrow 0$, the low $z$ behavior is linear
for hull tethering while it seems to be faster than linear
for all boundary tethering.  Also while the hull tethering scaling function
tends to a finite limit as $z \rightarrow \infty$ (see above discussion),
that for all boundary tethering appears to grow unbounded as it
should if $x<1$.
Corresponding results for all boundary tethering for the simple cubic lattice
are shown in Fig.4. Here the reasonable data collapsing is achieved for
$x=0.4$.

We find, in particular, that hull tethering cannot change the behavior
of the lowest energy mode, which remains the same as that of the untethered
case ($\epsilon_1 \sim S^{-d_w/d_f}$). This is because a fully tethered hull
confines the mode in the interior of the critical percolation cluster
which, asymptotically for large clusters, scales with the same fractal
dimension as the entire cluster itself. Thus, the lowest energy mode
results in the vibration of the interior of the cluster independent of the
external boundary. This inability to change the behavior of the lowest
energy mode is analogous to the case of an Euclidean cluster as well as
to the fractal drums (fractal external boundary and Euclidean interior)
\cite{sapoval}).

In contrast, the tethering of all boundaries leads to confinement
of the lowest energy vibrational mode in regions of the cluster
which are essentially compact, altering its behavior from a power-law
dependence on cluster size $\epsilon_1 \sim S^{-d_w/d_f}$
to a slow logarithmic dependence on $\epsilon_1 \sim (\ln S)^{-2/d_0 }$.
Thus all boundary tethering conspires with the complex geometry of the
critical percolation cluster (with pockets of compact regions, etc.)
to dramatically alter the behavior of lowest energy vibrational mode.

\section{Scaling of the maximum in density of states}
\label{peakscaling}

Let us now consider the vibrational density of states $\rho (\epsilon )$
of the matrix ${\bf W}$. For nontethered limit, $\rho (\epsilon )$
generally has a power law increase toward the lowest energy (stationary)
mode \cite{sonali}. On the other hand, if all boundary sites are tethered,
we recover the {\em ideal chain} result of a stretched exponential
{\em decrease} toward the lowest energy mode $\epsilon_1$ \cite{achille}.
Thus, when progressively more of all boundary sites are tethered, a
crossover between the two limits occurs. For an intermediate fraction $f$
of tethered sites, we generally observe a maximum in the density of states as
shown in Fig.5(a) for the square lattice. Starting from larger $\epsilon$
and moving toward $\epsilon =0$, $\rho$ increases initially because the
nontethered modes for the range of $\epsilon$ are too localized to be
affected when tethers are added, while for much smaller $\epsilon$ the tethers
begin to drag down the density of states. (Remember that tethers correspond
to traps and low $\epsilon$ corresponds to the long time survival of a
diffusing particle.) Thus a peak in $\rho (\epsilon )$
occurs, say, at $\epsilon_p$, which gradually increases from 0 for $f=0$
({\em ant} limit) toward $f=1$ (ideal chain limit) as the tethering fraction
$f$ increases.

The existence of the maximum in the density of states has implications
on the possible resonant behavior in response to the external source of
vibrational energy. See Leibig \cite{leibig} for a discussion of this
point for the case of weakly disordered network of Hookian springs.

For the hull tethering case, the situation is similar for finite $S$. In the
asymptotic large $S$ limit, however, hull tethering cannot affect the density
of states in qualitative manner no matter how large the tethering fraction
$f$ may be because the hull becomes increasingly a negligible fraction of
all boundaries. This implies that, for large $S$, the region near
$\epsilon =0$ must always witness power-law increasing $\rho (\epsilon )$.
However, the finite $S$ effect often masks this asymptotic behavior so that
for most sizes of clusters numerically generated and for most tether
fraction $f$, we do observe a maximum in the density of states similar to
the all boundary tethering case (see Fig.5(b)).

The crossover of $\epsilon_p$ as the tether fraction $f$ is varied is
then expected to obey the same type of scaling as for the lowest energy mode
$\epsilon_1$. For hull tethering,
\begin{equation}
\epsilon_p \sim S^{-d_w/d_f} G_p ( N ) ,
\label{eq:hullpeak}
\end{equation}
while, for all boundary tethering,
\begin{equation}
\epsilon_p \sim S^{-d_w/d_f} \overline{G}_p ( N/S^x ) ,
\label{eq:allpeak}
\end{equation}
with the same value of $x$ as in Eq.(\ref{eq:maxall}).

These scaling laws have been tested on the same critical percolation 
clusters as for the lowest energy mode scaling. The numerical results,
shown in Fig.6 for the hull case and Fig.7 for all boundary tethering,
are in good agreement with the scaling laws above. The corresponding results
for the simple cubic lattice in three dimensions are shown in Fig.8,
again with reasonably good agreement when a choice of $x=0.4$ is used.

\section{Spatial extent of the lowest energy mode}
\label{spatial}

The difference between hull and all boundary tethering is
dramatic also in the spatial characters of the normal modes.
We show in Fig.9 and Fig.10 the amplitude maps of typical lowest energy
modes (corresponding to $\epsilon_1$) for hull tethering and all boundary
tethering, respectively. For hull tethering, starting from low tethering
fraction $f$ (or small $N$) and increasing $f$ initially reduces the
amplitude of vibration of the sites in the vicinity of the tethered hull
sites.  This causes a decrease in the wavelength and consequently an increase
in $\epsilon_1$ (also of the scaling function $F(z)$). However after 
a certain number of hull sites are tethered, the amplitudes of the mode
at all sites near the hull become attenuated and the region of large
amplitudes becomes well confined to the interior of the cluster.
Beyond this point, increasing $N$ does not further affect the spatial
structure of the normal mode as the sites with appreciable amplitudes 
are already deeply in the interior of the cluster. This saturation behavior
may be likened to herding of livestock into a safe, fenced haven, and thus
we may call it {\em herding} of normal modes. Herding is reflected in the
saturation behavior of $F(z)$ for $z$ tending to $infinity$. 
In the examples shown in Fig. 9, the participation rations indicate that
only about 1\% of the sites contribute substantially to the mode for all
shown values of the tethering fractions.

For all boundary tethering, increasing $f$ serves to progressively
confine the (originally most extended, untethered) normal mode to sites
which are interior to both the external and internal boundaries, i.e.,
in the {\em blobs} of the percolation cluster.  Since typically most
blobs are very small, this results in a much sharper decrease in the
spatial extent of the normal mode, as seen in Fig.10.
Even for relatively small $f$, the effect of confinement can be nearly
complete in this case, and further tethering may force the mode to
jump around to distant locations which may allow the largest spatial
extent. This behavior may be likened to chasing wild animals in a
hunting expedition, and thus we may call it {\em hunting} of normal modes.
It is interesting to note that, while the largest modes without tethering
are those of small $\epsilon$, they are also the ones most affected
by tethering (particularly all boundary tethering), and thus after
some tethers have been put in place, these modes are not necessarily
the spatially largest ones any longer. Indeed those modes shown in
Fig.10 have main contributions only from less than 0.1\% of the sites
according to the values of the participation ratios.

The above observations on the spatial structure of normal modes under
boundary tethering may have technological implications. For example,
a soft glassy material may need to be clamped only at relatively few
locations on its external boundary to fully confine its lowest energy
normal mode to its interior via {\em herding}.

\section{Conclusion}
\label{conclusion}

The present work reveals a number of interesting features of the
elastic properties of a fractal with tethered boundaries. 
The vibrational modes and consequently the vibrational 
density of states for both hull and all boundary tethering
are dictated by regions of clusters which are {\em tether free}. Despite    
the structural similarity between the hull and internal boundaries,
the effect of tethering depends greatly on whether only the hull is
tethered or all boundaries are tethered. 
Tethering of hull sites confines the low-energy vibrational
modes in the interior of the cluster and since the interior
of the cluster is itself a fractal with the fractal dimension
same as that of the entire cluster, the leading behavior of the
vibrational density of states for fully tethered hull must
asymptotically be the same as that of the untethered cluster.
On the other hand, for full tethering of all boundaries,
the low-energy modes are confined in the compact regions of the cluster
which, because of their small spatial extent, give rise to  
faster than exponential decrease in the density of states
in the same limit.

Keeping in mind the fact that Euclidean clamped boundaries do not
have an appreciable effect on the density of states, our results
demonstrate the potential of tethering fractal boundaries to attenuate
harmonic excitations. Also, the present problem is equivalent to that
of diffusion in the presence of permanent traps, a fact that was taken
advantage of for the numerical part of this work. Since the latter
serves as a model for diffusion controlled reaction/absorption in the
presence of immobile reactants/absorbents, there may also be
applications in the areas of drug reaction/absorption, etc. Finally,
because of the similar mathematical formulation of quantum mechanical
localization and hopping transport problems \cite{odagaki}, we
expect that the same techniques will be useful in studying those
problems and that some of the current results may have direct
analogs in them.

\newpage
\begin{figure}
\caption{The average distance $l$ between tethered sites, average
diameter of the cluster $L_c$, and the average {\em blob} diameter $L_b$,
for (a) all boundary tethering and (b) hull tethering.}
\label{fig1}
\end{figure}

\begin{figure}
\caption{Scaling of the lowest energy mode for hull tethering on
square lattice. Scaling variable reduces to $N$, which is the total
number of tethered hull sites.}
\label{fig2}
\end{figure}

\begin{figure}
\caption{Scaling of the lowest energy mode for all boundary tethering on
square lattice. The scaling variable is $N/S^x$ where $x$ is numerically
about 0.54.}
\label{fig3}
\end{figure}

\begin{figure}
\caption{Scaling of the lowest energy mode for all boundary tethering on
simple cubic lattice. The scaling variable is $N/S^x$ where $x$ is numerically
about 0.4.}
\label{fig4}
\end{figure}

\begin{figure}
\caption{(a) Density of energy eigenvalues are shown for all boundary 
tethering of 5000-site clusters at the tethering fraction of 0.1, 0.4, and 
0.6. (b) shows the same for hull only tethering at fraction 0.3, 0.5, and 0.9.}
\label{fig5}
\end{figure}

\begin{figure}
\caption{Scaling of the peak in the density of states for hull tethering
on square lattice.}
\label{fig6}
\end{figure}

\begin{figure}
\caption{Scaling of the peak in the density of states for all boundary
tethering on square lattice.}
\label{fig7}
\end{figure}

\begin{figure}
\caption{Scaling of the peak in the density of states for all boundary
tethering on simple cubic lattice.}
\label{fig8}
\end{figure}

\begin{figure}
\caption{Spatial extent of the lowest energy mode is illustrated by plotting
the amplitudes of the mode for a 50000 site cluster on the square lattice 
for progressively increasing hull tethering fraction. Grey scale is used 
where lightest grey is used for the lowest (absolute value of) amplitude
and black for the highest, in logarithmic scale. (a) corresponds to
random tethering of 20\% (upper left), (b) to 40\% (upper right),
(c) to 70\% (lower left), and to 100\% (lower right). Note that with zero
tethering, the amplitude will be completely uniform.}
\label{fig9}
\end{figure}

\begin{figure}
\caption{Same as Fig.9 but for the all boundary tethering case.}
\label{fig10}
\end{figure}

\end{document}